\documentclass[sn-standardnature,iicol]{sn-jnl}

\usepackage[compact]{titlesec}
\titleformat{\section}{\normalfont\normalsize\bfseries}{0pt}{}{}
\usepackage{siunitx}
\usepackage{gensymb}
\usepackage{anyfontsize}

\jyear{2024}%

\begin{document}

\title[Article Title]{Coexisting phases of individual VO$_2$ nanoparticles for multilevel nanoscale memory}

\author*[1,2]{\fnm{Peter} \sur{Kepič}}\email{peter.kepic@ceitec.vutbr.cz}

\author[2]{\fnm{Michal} \sur{Horák}}\email{michal.horak2@ceitec.vutbr.cz}

\author[1,2]{\fnm{Jiří} \sur{Kabát}}\email{jiri.kabat@vutbr.cz}

\author[1,2]{\fnm{Vlastimil} \sur{Křápek}}\email{krapek@vutbr.cz}

\author[1,2]{\fnm{Andrea} \sur{Konečná}}\email{andrea.konecna@vutbr.cz}

\author[1,2]{\fnm{Tomáš} \sur{Šikola}}\email{tomas.sikola@ceitec.vutbr.cz}

\author*[1,2]{\fnm{Filip} \sur{Ligmajer}}\email{filip.ligmajer@ceitec.vutbr.cz}

\affil[1]{\orgdiv{Central European Institute of Technology}, \orgname{Brno University of Technology}, \orgaddress{\street{Purkyňova 123}, \city{Brno}, \postcode{612 00}, \country{Czech Republic}}}

\affil[2]{\orgdiv{Institute of Physical Engineering, Faculty of Mechanical Engineering}, \orgname{Brno University of Technology}, \orgaddress{\street{Technická 2}, \city{Brno}, \postcode{616 69}, \country{Czech Republic}}}

\abstract{Vanadium dioxide (VO$_2$) has received significant interest in the context of nanophotonic metamaterials and memories owing to its reversible insulator-metal transition associated with significant changes in its optical and electronic properties. While the VO$_2$ transition has been extensively studied for several decades, the hysteresis dynamics of individual single-crystal VO$_2$ nanoparticles (NPs) remains largely unexplored. Here, employing transmission electron microscopy techniques, we investigate phase transitions of single VO$_2$ NPs in real time. Our analysis reveals the statistical distribution of the transition temperature and steepness and how they differ during forward (heating) and backward (cooling) transitions. We assess the stability of coexisting phases in individual NPs and prove the persistent multilevel memory at near-room temperatures using only a few VO$_2$ NPs. Our findings shed new light on the underlying physical mechanisms governing the hysteresis of VO$_2$ and establish VO$_2$ NPs as a promising component of optoelectronic and memory devices with enhanced functionalities.}


\maketitle

\begin{figure*}
    \centering
    \includegraphics{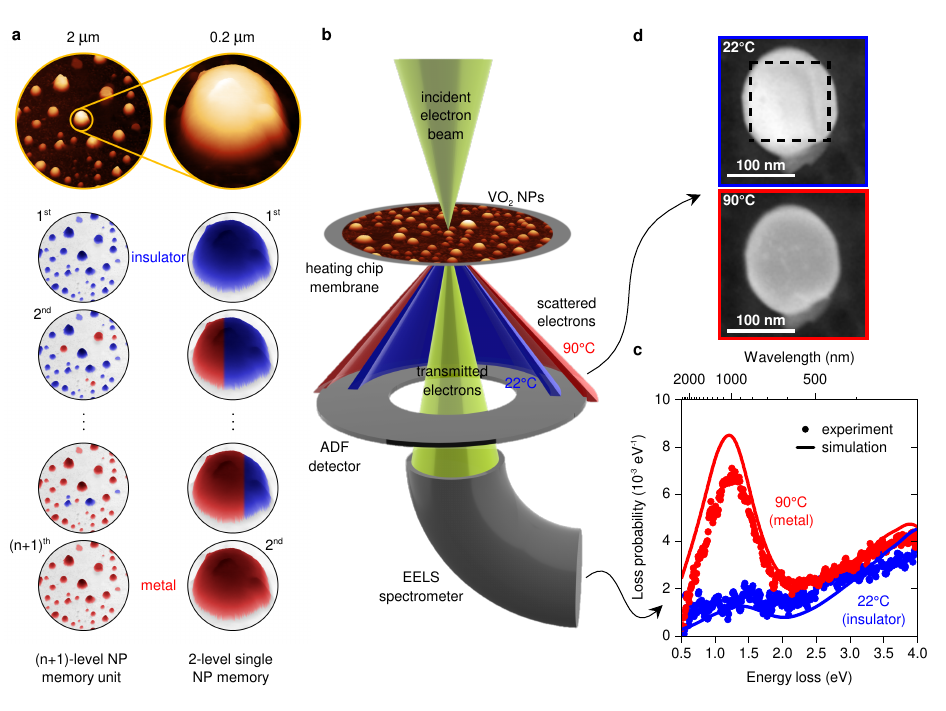}
    \caption{\textbf{VO$_2$ NP memory and the measurement setup.} \textbf{a}, Concept of a VO$_2$ NP memory unit based on coexisting insulator (blue) and metal (red) phases within a single VO$_2$ NP. The left column corresponds to (n+1)-level memory based on an n-NP ensemble, and the right column corresponds to a 2-level single-NP memory. The illustration uses real atomic force microscopy (AFM) images of our samples. \textbf{b}, Scheme of EEL spectroscopy and ADF measurement of VO$_2$ NPs at different temperatures. \textbf{c}, Measured and simulated EEL spectra of a \SI{130}{nm} VO$_2$ NP on a SiN membrane in the insulator (\SI{22}{\celsius}; blue) and metal (\SI{90}{\celsius}; red) phases. \textbf{d}, ADF images of the NP in (\textbf{c}) at the listed temperatures. The dashed rectangle highlights the area over which EEL spectra in (c) were spatially integrated.}
    \label{fig1}
\end{figure*}

Phase-change materials, with their solid--solid transition accompanied by a significant change of the optical properties, play an important role in the rapidly developing field of nanophotonics. Tunable metasurfaces for active control of light waves~\cite{Yang20221} or integrated photonic memories for novel types of computing and data storage~\cite{Lian2022} represent key areas of these materials applications. On-and-off states of the optical memories are represented by high and low optical transmission, which is controlled by materials that have multiple phases with distinct optical properties~\cite{Youngblood2023}. The most utilized materials in this field are chalcogenide glasses, typically germanium antimony telluride (GST) compounds, as they exhibit a non-volatile amorphous-crystalline phase transition accompanied by significant optical modulation~\cite{Wuttig2017}. For example, a patch of GST on top of a waveguide can store several levels of optically readable information and retain them for decades~\cite{Rios2015}. Unfortunately, while energy consumption for retaining the data is zero, writing them (amorphizing the material) requires an ultrafast melt-quench process at above \SI{600}{\celsius} and \SI{10}{\celsius/ns}~\cite{Abdollahramezani2022}. This process can be very energy-consuming, especially when only volatile memory is sufficient, as in neuromorphic computing~\cite{Schofield2023}, random access memories~\cite{Lee2024}, or reconfigurable metasurfaces~\cite{Gu2023}. In such scenarios, a volatile phase-change material, vanadium dioxide (VO$_2$), represents a more efficient solution because its insulator-metal transition (IMT) is available at around \SI{68}{\celsius} and at microscale requires only nanojoules of energy~\cite{Parra2021, Jung2022, Jung2021}. Single-crystal VO$_2$ nanoparticles (NPs), with their reverse metal-insulator transition (MIT) reaching \SI{30}{\celsius}~\cite{Lopez2002, Nishikawa2023}, could lower the retention energy even more by bringing the whole platform closer to the room temperature. Moreover, they could increase the bit compactness (the number of optical levels per area) because the exact transition temperature and absorption of each (lithographically) separated NP (see Fig.~\ref{fig1}a) is easier to control than that of a film where the transition is more stochastic~\cite{Cheng2021}, and the absorption is averaged out. The coexistence of phases, already demonstrated for single crystal films~\cite{Johnson2022, Qazilbash2007} and nanobeams~\cite{Huber2016, Bae2013}, brings another degree of freedom that can be exploited at the NP level for multilevel memories.

So far, the hysteresis characteristics (transition temperature, steepness, and contrast) and their control by stoichiometry, substrate-induced strain, volume, defects and doping have been studied for VO$_2$ films~\cite{Andrews2019, White2021, Wan2019}, and ensembles of VO$_2$ NPs~\cite{Lopez2002, Donev2009, Appavoo2012, Clarke2018, Nishikawa2023}. However, the hysteresis characteristics of individual single-crystal VO$_2$ NPs remain elusive, although such NPs represent the smallest unit (monocrystalline grains) of the polycrystalline films and large nanostructures. The possible coexistence of the insulating and metallic phases, observed so far only in microscale nanobeams~\cite{Lei2015, Zhi2014}, could bring another degree of freedom to multilevel memories. Moreover, the low-temperature stability of the coexisting phase and the mechanism of their formation are yet to be examined at the level of individual NPs.

Here, we report on our study of individual single-crystal VO$_2$ NPs by scanning transmission electron microscopy (STEM) techniques, which provide superior spatial resolution and correlation of high-resolution images with the spectroscopic information~\cite{Yang20222, Hage2020}. While we thermally induced the phase transition, we correlated the evolution of electron energy loss (EEL) spectra with annular dark-field (ADF) structural contrast (see Fig.~\ref{fig1}b). Thus, we have been able to recognize the individual coexisting phases and analyze their spatial distribution and hysteretic behaviour within individual VO$_2$ NPs in real time. By investigating the hystereses of hundreds of NPs, we have found that the macroscale VO$_2$ IMT consists of gradual transitions with transition temperatures narrowly spread, while the MIT is formed by extremely abrupt transitions at temperatures that are spread across a larger temperature window. Ultimately, we examined the stability of the coexisting phases at lower temperatures for both individual NPs and NP ensembles. Based on these findings, we demonstrate an optical memory unit in the form of a VO$_2$ NP ensemble, where the number of memory levels relates to the number of NPs (see Fig.~\ref{fig1}a). These results add on new possibilities to the already established VO$_2$ optical memory devices~\cite{Jung2022, Jung2021}, and strengthen the position of VO$_2$ as an optical unit for information storage and processing.

\begin{figure*}
    \centering
    \includegraphics{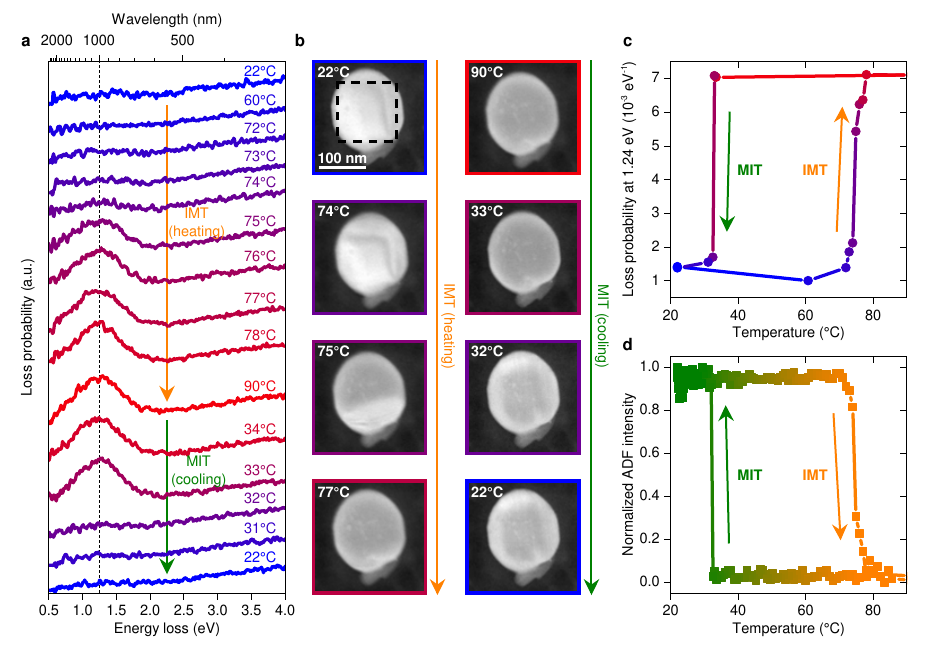}
    \caption{\textbf{Phase transition hysteresis of a single VO$_2$ NP.} \textbf{a}, EEL spectra of a single VO$_2$ NP recorded at temperatures listed in the graph, as the NP is driven through its IMT and back. The vertical dashed line highlights the position of the plasmon resonance, and the arrows indicate the direction of the IMT (orange, heating) and MIT (green, cooling). \textbf{b}, ADF images of the NP at the listed temperatures. The dashed rectangle highlights the area over which EEL spectra in (\textbf{a}) were spatially integrated. \textbf{c}, Phase transition hysteresis extracted from the EEL spectra in (\textbf{a}), averaged between 1.0--\SI{1.5}{eV}. \textbf{d},  Phase transition hysteresis extracted from the mean ADF intensity of the NP in (\textbf{b}).}
    \label{fig2}
\end{figure*}

\section*{Probing the phase transition by STEM}

To verify the possibility of probing phase transitions of individual VO$_2$ NPs by STEM, we fabricated them by dewetting a VO$_2$ film (see Methods) on a commercial SiN membrane on a heating chip (Protochips Inc.). The dewetting process was chosen because it provides well-separated single-crystal hemispherical NPs (see images and size distribution in Extended Data Fig.~\ref{extfig1}), with low MIT temperatures even on top of amorphous substrates~\cite{Lopez2002, Nishikawa2023}.

Fig.~\ref{fig1}c shows the measured and simulated EEL spectra of a VO$_2$ NP (diameter \SI{130}{nm}), below and above the transition temperature. A prominent plasmonic peak, emerging at \SI{1.24}{eV} in the spectrum of the metallic phase (further described in Extended Data Fig.~\ref{extfig2}), clearly confirms the IMT. Investigating such transitions in other NPs (see Extended Data Fig.~\ref{extfig3}a), we found that the EEL contrast (the signal difference between the insulator and metal phase) linearly increases with the NP diameter. As this trend relates to the increasing optical absorption of NPs (see Extended Data Fig.~\ref{extfig3}b), the results from the EEL measurement can be directly linked to the far-field optical properties of VO$_2$ NPs. 

When looking at the NP image obtained by the ADF detector (Fig.~\ref{fig1}d), we can see a distinct change in the contrast upon the IMT. As the ADF detector records mostly elastically scattered electrons at a specific angular range~\cite{Nellist2000}, the change of ADF contrast can be related to the structural transformation during the transition of the NP. We verified this connection in a recent comprehensive study, where various TEM and STEM techniques were used to characterize lattice and electronic signatures of the phase transition (see \cite{Krapek2024}). With EEL spectroscopy and ADF at our disposal, we can extend our analysis toward real-time phase transition dynamics at the level of individual NPs, which would be otherwise hardly accessible using far-field optical measurements.

\begin{figure*}
    \centering
    \includegraphics{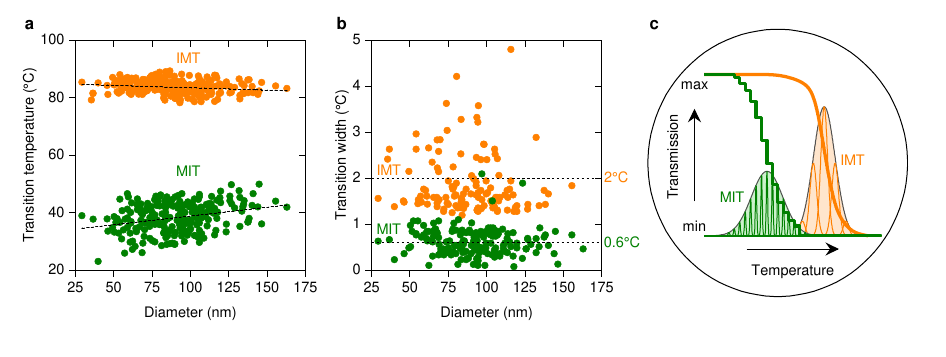}
    \caption{\textbf{Statistics of the phase transition hysteresis of VO$_2$ NPs.} \textbf{a}, Size-dependent distribution of transition temperatures and \textbf{b}, transition widths of the IMT (orange) and MIT (green) of various NPs. Dashed lines represent linear fits and averaged values in (\textbf{a}) and (\textbf{b}), respectively. \textbf{c}, Schematic of the phase transition hysteresis of a polycrystalline VO$_2$ film, which we model as a system of individual NPs. Each NP is represented by a peak, with the peak widths and positions reflecting the statistical distributions obtained in panels \textbf{a} and \textbf{b}. The overall behaviour (e.g. transmission) of the polycrystalline film is then an aggregate of transitions of the individual grains (modelled by NPs).}
    \label{fig3}
\end{figure*}

\section*{Hysteresis of individual VO$_2$ NPs}

The EEL spectra and ADF images of the NP investigated in Fig.~\ref{fig1}c, d were sequentially taken during the phase transition to observe its dynamics. In Fig.~\ref{fig2}a, we can see the temperature evolution of the EEL spectra during the heating and cooling cycle. Upon extracting the EEL probability (averaged between 1.0--\SI{1.5}{eV}) and plotting it as a function of temperature, we can observe the single-NP hysteresis (Fig.~\ref{fig2}c). The gradual IMT around \SI{74.5}{\celsius} indicates the presence of coexisting phases (further confirmed in Extended Data Fig.~\ref{extfig4}). The abrupt MIT at \SI{32.7}{\celsius}, on the other hand, indicates an avalanche-like transition from a supercooled state~\cite{Sharoni2008, Uhlir2016}. Note that such a transition does not exclude the presence of coexisting phases during the MIT, but it makes their observation harder: the activation volume (the minimum size of an initial nucleus) for the MIT is often greater than that for the IMT~\cite{Motyckova2023}, and can be even larger than the volume of the investigated NP~\cite{Jin2022}. Note that we observed this phase coexistence also in other NPs (see Extended Data Fig.~\ref{extfig5}b). Interestingly, the phase coexistence was also visible in ADF images, as shown in Fig.~\ref{fig2}b, where the NP gets darker upon the IMT. By processing the sequence of ADF images (see Methods), we were able to reconstruct (Fig.~\ref{fig2}d) the same hysteresis of the NP as with the EEL data. Overall, the ADF approach was three times faster and delivered a hundred times lower electron dose to the sample than the EEL measurement, which is especially important when one wants to avoid the spurious influence of the high-energy electrons on the sample (see Extended Data Fig.~\ref{extfig5}). The ADF technique thus facilitates localized analysis of structural transitions and has the potential for examining other phase-change materials at the nanoscale or disentangling the mechanism of non-volatile ramp-reversal memory behavior of VO$_2$~\cite{Fried2024, Basak2023}. In our case, it allowed us to analyze the phase transition hysteresis of hundreds of VO$_2$ NPs simultaneously and to discover important differences between the IMT and MIT, as we will describe now. 

\section*{Statistics of VO$_2$ hysteresis properties}

In most polycrystalline VO$_2$ films and nanostructures, the steepness of the IMT is practically identical to that of the MIT~\cite{Cueff2020, Appavoo2012, Kepic2021, Ligmajer2018}. There are, however, films where these two transitions differ significantly~\cite{Suh2004, Kovacs2011}. To better understand the shape of these large-scale transitions from the single-grain perspective, we recorded video-rate sequences of ADF images (\SI{2}{fps}, \SI{1}{\celsius/s}, see Extended Video 1) and utilized them to simultaneously analyze phase transition hystereses of hundreds of VO$_2$ NPs. The extracted size-dependence of transition temperatures during the IMT (orange) and MIT (green) of 266~VO$_2$ NPs is plotted in Fig.~\ref{fig3}a. While the IMT temperature is weakly dependent on the NP diameter (decreases by \SI{0.02}{\celsius/nm}), the MIT temperature is three times more sensitive to the increasing diameter (increases by \SI{0.06}{\celsius/nm}). The hysteresis broadening towards the lower NP volume observed here and by others~\cite{Lopez2002, Clarke2018} thus seems to be predominantly related to the MIT. Moreover, the statistical variation across all NPs, which arises from the local inhomogeneities within individual NPs~\cite{Huber2016} and the inherent stochasticity of the transitions~\cite{Basak2023, Cheng2021}, is more than twice as wide for MIT temperatures ($\Delta T \approx$ \SI{24}{\celsius}) compared to IMT temperatures ($\Delta T \approx$ \SI{11}{\celsius}). 

We also analyzed transition widths of our NPs (the temperature range it takes to switch a NP fully from one phase to another) and report the results in Fig.~\ref{fig3}b: While we cannot see any clear size dependence of the IMT and MIT transition widths, we find that the average IMT width (\SI{2.0}{\celsius}) is more than three times larger than the average MIT width (\SI{0.6}{\celsius}). Note that such a narrow MIT width can be even an overestimation, as it was limited by the chosen \SI{0.5}{\celsius} step and by the curve-fitting process (see Methods). Despite the very abrupt nature of the MIT in the vast majority of NPs, three of them (\SI{1}{\percent}) exhibited coexisting phases during the MIT nevertheless (see the green data points above \SI{1}{\celsius} in Fig.~\ref{fig3}b and Extended Data Fig.~\ref{extfig6}). Our findings confirm that the co-existence of insulating and metallic phases in VO$_2$ of microscale volumes~\cite{Huber2016, Bae2013} is also present at the previously unexplored limit of nanoscale volumes. Moreover, the coexisting phases at the nanoscale are at least 70-times less probable during the MIT than during the IMT (3 NPs versus 210~NPs). The observed difference in the transition widths indirectly confirms that the activation volume during the MIT (discussed in the previous section) is larger than during the IMT. In the context of a hypothetical multilevel phase-coexistence memory based on VO$_2$ NPs, the retention temperature can thus be set right above the abrupt MIT, which is near the ambient temperature.

\begin{figure*}
    \centering
    \includegraphics{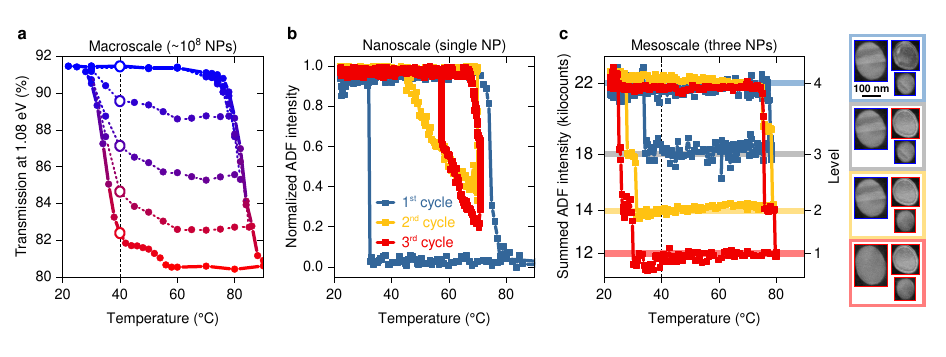}
    \caption{\textbf{Stability of coexisting phases and a mesoscale optical VO$_2$ memory unit.} \textbf{a}, Phase transition hysteresis of a macroscale ensemble of VO$_2$ NPs extracted from their far-field transmission at \SI{1.08}{eV}, with one full and three partial heating--cooling cycles. A multilevel far-field optical memory can be realized with such an ensemble at a temperature as low as \SI{40}{\celsius} (vertical dashed line). Note the memory levels are well separated from each other, beyond the noise observable at the outermost hysteresis \textbf{b}, Stability of the coexisting phases within a single nanoscale VO$_2$ NP, examined during one full and two partial heating--cooling cycles using normalized ADF intensity. The unstable coexistence of phases leads to spontaneous MIT. \textbf{c}, Summed ADF intensity of three VO$_2$ NPs (a mesoscale memory concept) as they experience three heating--cooling cycles. The data indicate that such an ensemble can represent a 4-level mesoscale memory stable at \SI{40}{\celsius}. Micrographs next to the graph show ADF images of the studied NPs during those three cycles, with blue and red rectangles highlighting the insulating and metallic phases, respectively. Note that NP images are artificially assembled to be next to each other compared to the original positions in the video.}
    \label{fig4}
\end{figure*}

Based on these statistical findings, we put forward a model for the phase transition of polycrystalline VO$_2$ films and microstructures, where the size of grains is generally similar to the sizes of NPs investigated here. So far, the phase transition was modeled based on the macroscopic properties of such films or on aggregated properties of discrete NP ensembles~\cite{Wan2019, Lopez2002, Nishikawa2023}. In our description, the macroscopic forward transition is composed of the IMTs of individual grains, which have broad transition widths, but their transition temperatures have a very narrow spread around the central value (Fig.~\ref{fig3}c). During the backward MIT, on the other hand, the transitions of the grains are very abrupt (narrow MIT widths) but are spread across a larger temperature window ($\Delta T$), which we justify in our model by inequalities of grain boundaries~\cite{Appavoo2012} and strain~\cite{Huber2016} in a polycrystalline film. These findings shed new light on the physics of VO$_2$ hysteresis and coexisting phases within the NPs, which are important for designing optical memories. In the following paragraphs, we will focus more on the concept of a multilevel optical memory based on VO$_2$ NPs and further investigate the low-temperature stability and hysteresis of the individual NP coexisting phases.

\section*{Multilevel optical memory}

Already in 2015, Lei et al.~\cite{Lei2015} demonstrated a persistent memory effect in a hybrid system composed of a VO$_2$ substrate and plasmonic nanoantennas. In their work, multiple memory levels of optical extinction spanning the full range of the hysteresis loop were optically addressed when the hybrid system was latched at different starting temperatures during the IMT. Analogically, we latched the VO$_2$ NPs in various mixtures of phases by terminating the IMT at distinct temperatures. This allowed us to create a macroscale free-space optical memory device (Fig.~\ref{fig4}a) that possesses multiple levels of transmission states even at a convenient temperature of \SI{40}{\celsius}. However, to reduce the footprint and scale down a potential memory device, it is essential to understand if the intermediate transmission levels are formed by a mixture of fully insulating and fully metallic NPs or by NPs that are switched to different degrees, i.e., with coexisting phases.

Our ADF imaging strategy enabled us to investigate the stability and coexistence of VO$_2$ phases within individual NPs. We focused on the NP already shown in Fig.~\ref{fig2}, which exhibited such a coexistence and applied three heating--cooling cycles with the heating terminated at three distinct temperatures to stabilize three levels of the phase coexistence (Fig.~\ref{fig4}b). In the first cycle, the full IMT was completed, and the heating terminated at the temperature of \SI{90}{\celsius}. The NP then retained its metallic state upon cooling by more than \SI{50}{\celsius} down to \SI{32}{\celsius}. In the second cooling cycle, when the heating terminated at \SI{70}{\celsius} while the NP exhibited a phase coexistence (with approximately \SI{65}{\percent} of the metallic phase), the NP was continuously switching back to the insulating state instead of partially retaining the metallic state, and the MIT was completed already at \SI{45}{\celsius}. In the third cycle, when we terminated the heating at \SI{71}{\celsius} (with approximately \SI{80}{\percent} metallic phase) and cooled the system down to \SI{57}{\celsius}, we even observed spontaneous propagation of the insulating phase at the constant temperature by approximately \SI{0.2}{nm/s} till the NP was fully switched. Based on these observations, also confirmed by several other NPs (see Extended Video 2), we infer that coexisting phases within single VO$_2$ NPs are not stable at lower temperatures, which means there are no hysteresis sub-loops analogical to those of macroscopic NP ensembles (Fig.~\ref{fig4}a). However, the slowly moving phase-front does not preclude the coexistence of phases at shorter time scales. To identify the true limits of coexisting phases within the MIT of VO$_2$ NPs for memory and neuromorphic computing applications~\cite{Schofield2023}, a rigorous ultrafast study is needed, in analogy to studies done on VO$_2$ films, single-crystalline nanobeams and NP ensembles~\cite{Huber2016, Johnson2022, Lopez2004}. We envision that the analogous image contrast changes can help to resolve this question when exploited in an ultrafast pump-probe TEM setup~\cite{Danz2021}. Until then, the unstable metallic phase of the individual NP at low temperatures conclusively confirms that the intermediate transmission levels of large-scale devices (such as in Fig.~\ref{fig4}a) are formed by a mixture of fully insulating and fully metallic NPs (as proposed in~\cite{Lopez2002} and recently indicated in~\cite{Nishikawa2023}).

Although a macroscopic ensemble of VO$_2$ NPs can be used as a persistent multilevel memory, the low stability of coexisting phases in single individual VO$_2$ NPs prevents their use in nanoscale long-term multilevel memory applications. These two extreme cases imply there must exist an intermediate design corresponding to a mesoscale ensemble of VO$_2$ NPs that can retain a delocalized intermediate state with the lowest possible footprint. We, therefore, experimentally analyzed phase transition hysteresis of an ensemble of three VO$_2$ NPs using the metric of their collective ADF intensity. We chose those particular NPs because their footprint can be kept below \SI{500}{nm} $\times$ \SI{500}{nm} and because they exhibited sequential, cumulative switching during at least three recorded cycles (see Extended Video 2). In Fig.~\ref{fig4}c, we can see the sum of ADF intensity of the three VO$_2$ NPs. During the first cycle, where heating stopped at \SI{78}{\celsius}, only one NP went through the IMT while maintaining its metallic phase till \SI{34}{\celsius}. During the second cycle (stopped at \SI{79}{\celsius}), also the second NP was switched, albeit before the first NP. This inconsistency with the first cycle, which arises from the statistical distribution of the IMT temperatures, must be properly addressed when the actual integrated optical memory is designed. The reverse MIT during the second cycle occurred below \SI{31}{\celsius} for both NPs. During the third cycle (stopped at \SI{80}{\celsius}), also the last NP finally switched into the metallic state, and all NPs returned to their insulating states when cooled below \SI{30}{\celsius}. Note that the MIT temperature of the first two NPs decreased after each cycle. Such a sequential decrease most likely results from the accumulated electron dose, which was shown to drive the NPs to ultimately retain the metallic state even at room temperature (Extended Data Fig.~\ref{extfig5}). Although unwanted during analysis like this one, such an effect can be useful for post-fabrication trimming of VO$_2$ hysteresis.

In the previous sections, we showed that the change in ADF intensity during the phase transition relates to the change in the EEL spectrum and that EEL properties can be directly linked to optical properties (see Extended Data Fig.~\ref{extfig3}b). It means that ADF levels observed in Fig.~\ref{fig4}c can be considered as four optical transmission levels, e.g. when NPs are placed on top of a waveguide. Such an ensemble of $n$ NPs thus represents a mesoscale version of the memory unit described at the beginning of this section and shown in Fig.~\ref{fig4}a, where the memory levels scale as $n+1$. Because the MIT of these NPs occurred below \SI{40}{\celsius}, the retention energy of such memories is significantly lower than that of the current VO$_2$ optical memories. However, for larger NPs, where transmission contrast is larger (as shown in Extended Data Fig.~\ref{extfig3}a), the hysteresis is narrower and thus offers less room for memory stability. As there might be a certain minimal required transmission change for a practical memory application, this contrast--width trade-off must be carefully considered during the design process. Another consideration must be taken when choosing and addressing NPs due to the statistical distribution of the IMT temperatures. Nevertheless, the demonstrated VO$_2$ NP memory, which provides several levels of optical transmission with a sub-micron footprint accessible by ultra-fast electrical or optical pulses with low peak powers, proves to be another noteworthy candidate for optical memories used in future data storage and processing.

\section*{Conclusion}

In conclusion, we have studied individual single-crystal VO$_2$ NPs with STEM techniques and showed that their plasmonic fingerprints, appearing in the EEL spectra of NPs in the metallic phase, serve as a reliable high-contrast tool for tracking the phase transitions. By correlating EEL measurements with far-field optical properties, we proved that larger NPs exhibit greater optical (absorption) contrast in the telecom region upon the transition. Then, we dynamically recorded the evolution of EEL spectra and ADF images during respective heating and cooling cycles. We obtained hystereses of individual NPs, verified the coexistence of the insulator and metal phases during the IMT and proved that the intensity of ADF images of NPs can be used to correctly recreate the hysteresis obtained by the EEL measurement. While the electron beam was found to gradually lower the MIT temperature and therefore impact the measurement, analyzing the transition using ADF images is a 100-times less-exposing and three-times faster technique than the EEL measurement. The ADF videos allowed us to simultaneously study 266~NPs and obtain their IMT and MIT temperatures and transition widths. We found that with the increasing NP volume, the MIT temperature also increases, and thus, the hysteresis becomes narrower, while the statistical distribution of the MIT temperatures is more than two times larger than that of the IMT. Based on these results and the evidence that the MIT is more than three times sharper than the IMT (coexisting phases 70-times less probable), we proposed a model for the overall hysteresis of a polycrystalline VO$_2$ film. In the model, the hysteresis is formed by IMTs of individual NPs, whose temperatures are spread narrowly but transition widths are broad, and by MITs, whose temperatures are distributed across a larger temperature window, but the transitions are extremely abrupt. Lastly, we showed that the coexisting phases of the individual NPs are not persistent at lower temperatures, and therefore, the individual NP cannot be used as a long-term multilevel memory kept at near-room temperature. This finding proved that the intermediate states (levels) presented in the macroscale transmission memory made of an ensemble of VO$_2$ NPs represent a "digital" mixture of either fully-switched or non-switched NPs. Finally, we established a concept of mesoscale memory working at mere \SI{40}{\celsius}, where the number of levels scales with the number of NPs as n+1. Despite the discussed contrast--width trade-off (larger NPs have more considerable transmission contrast but higher MIT temperature; larger retention temperature is needed), such VO$_2$ NP optical memory with already several levels of information at the sub-micron footprint can surpass current VO$_2$ optical memory patches. While unsuitable for long-term data storage due to the requirement for constant energy supply~\cite{Youngblood2023}, this low-temperature hysteresis with ultrafast transition makes VO$_2$ one of the candidates for optical random access memories~\cite{Lee2024} or for optical neuromorphic computing that often require only short-term data retention~\cite{Schofield2023}. 

\backmatter

\bmhead{Acknowledgments}
This work is supported by the Grant Agency of the Czech Republic (22-04859S) and by the project Quantum Materials for Applications in Sustainable Technologies (QM4ST, project No. CZ.02.01.01/00/22\textunderscore008/0004572) by OP JAK, call Excellent Research. We acknowledge CzechNanoLab Research Infrastructure supported by MEYS CR (LM2023051) for the financial support of the measurements and sample fabrication at CEITEC Nano Research Infrastructure. P.K. acknowledges support from the BUT specific research project, Brno PhD talent scholarship and Thermo Fisher Scientific scholarship. We thank J. A. Arregi and V. Uhlíř for fruitful discussions about phase-change materials.

\bmhead{Funding}
Grant Agency of the Czech Republic (22-04859S), MEYS CR (CZ.02.01.01/00/22\textunderscore008/0004572, LM2023051)

\bmhead{Competing interests}

The authors declare no competing interests.
\bmhead{Ethics approval} 

Not applicable
\bmhead{Consent to participate}

Not applicable
\bmhead{Consent for publication}

Not applicable
\bmhead{Availability of data and materials}

The datasets generated during and/or analyzed during this study are available from the corresponding authors upon reasonable request.

\bmhead{Code availability} 

Not applicable

\bmhead{Authors' contributions}

P.K. and M.H. conceived the idea, designed the study and fabricated the samples. P.K., M.H. and F.L. performed and evaluated measurements. P.K. and J.K. performed Ansys Lumerical and COMSOL Multiphysics simulations, respectively. V.K., A.K., T.S. and F.L. coordinated the research project. All authors contributed in writing and reviewing the paper.

\bibliography{sn-bibliography}

\begin{appendices}
\newpage

\section*{Methods}

\textbf{Fabrication}
VO$_2$ NPs on fused silica and SiN heating chip membrane (Protochips Inc.) were fabricated in a two-step process. First, the \SI{30}{nm} amorphous film was fabricated by TSST pulsed laser deposition (PLD) system (\SI{248}{nm} KrF laser, \SI{2}{J/cm^2}, \SI{10}{Hz}, \SI{50000}{pulses}, vanadium target (99.9\% purity, Mateck GmbH), \SI{50}{mm} substrate--target distance, room temperature and \SI{5}{mTorr} oxygen pressure) and then annealed ex-situ for \SI{30}{min} in a vacuum furnace (Clasic CZ Ltd.) at \SI{700}{\celsius} and \SI{15}{sccm} oxygen flow.

\textbf{Characterization}
Transmission and ellipsometry of the VO$_2$ NPs on fused silica and VO$_2$ film on silicon, respectively, were carried out using a spectroscopic ellipsometer J. A. Woollam V–VASE with an adjustable compensator, \SI{1000}{\micro m} monochromator slit, 50\degree, 60\degree \ and 70\degree \ incident angles (for ellipsometry), \SI{3}{mm} spot size and 0.5--\SI{4}{eV} spectral range with \SI{0.04}{eV} step. The extinction was calculated as 1 - T, where T is the transmission. The temperature was feedback-controlled during the spectroscopic measurements by a home-built heating stage, which included a resistive heater and a thermocouple.

Scanning transmission electron microscopy with EEL spectroscopy was performed using TEM FEI Titan equipped with a monochromator, GIF Quantum spectrometer for EELS, and in-situ Fusion Select system by Protochips for heating experiments. We used the following parameters: primary beam energy of \SI{120}{keV}, electron beam current around \SI{100}{pA}, convergence semi-angle of \SI{8.14}{mrad}, ADF collection angle 16.7--\SI{38.3}{mrad} and exposure time 1--\SI{2}{\micro s/px}, and EELS collection angle \SI{8.3}{mrad} and exposure time 0.2--\SI{0.3}{ms/px}. EEL spectra were integrated over marked regions of interest and further processed by background and zero-loss peak subtraction, normalized with respect to integral zero-loss peak intensity (energy window from -\SI{1}{eV} to +\SI{1}{eV}) to transform counts to a quantity proportional to the loss probability, and divided by the spectrometer dispersion \SI{0.01}{eV/px} to obtain loss probability density in units \SI{}{eV^{-1}} referred to as the loss probability. 

Note that the elevated temperature during the fabrication process corrupted the heating element on the membrane, so the heating chip temperatures had to be re-calibrated (see Extended Data Fig.~\ref{extfig7}).

\textbf{Processing of ADF intensity for hysteresis and statistics}
After recording the ADF video during the phase transition of VO$_2$ NPs using Protochips Inc. software Axon, we extracted a sequence of images and processed them using the ImageJ software~\cite{Schneider2012} and StackReg plugin~\cite{Thevenaz1998}. Specifically, we aligned images using the StackReg plugin, created a mask around each of the NPs, and measured the mean grey value of masked NPs simultaneously across all images/temperatures. The transition temperatures and transition widths of each NP were obtained from the position and full width at half maximum (FWHM) of a Gaussian profile fitted to the first-order derivative of either the IMT or MIT parts of the hysteresis curve. Transition widths were calculated as 6 standard deviations of the fitted normal Gaussian distribution, which equals 2.55 FWHM. This value is equivalent to the temperature range it takes to switch fully from one phase to another in a \SI{99.8}{\percent} confidence interval. In Fig.~\ref{fig4}c, the ADF intensities of NPs were summed (or subtracted in the case of the first switched NP) for a clear demonstration of the memory effect. In the case of an actual optical device, the transmitted intensity is less noisy and always drops during the IMT as the light absorption of the NP in the metallic phase is larger.

\textbf{Simulation}
Extinction of hemispherical VO$_2$ NP arrays on fused silica and extinction coefficient of single hemispherical VO$_2$ NPs in air were calculated using the finite-difference time-domain (FDTD) method implemented in the Ansys Lumerical FDTD Solutions software. The distance between the NPs and the horizontal FDTD region boundary was always kept at least half of the longest recorded wavelength. Conformal meshing (mesh order 4) was adopted everywhere except the NPs, where we set the staircase meshing with a \SI{5}{nm} step. We employed a total-field scattered-field (TFSF) source with 16 layers of perfectly matched layers boundary conditions for individual NP simulations and a plane wave source with periodic boundary conditions and double-diameter square spacing for the simulations of the arrays. The NPs were illuminated from the bottom/substrate. The extinction of arrays was calculated as 1 - T, where T is the transmission, which was obtained from the monitor above the NPs. The extinction coefficient of individual NPs was obtained by dividing the sum of scattering and absorption cross-sections by the diameter of the NP. All simulated and experimental data in Extended Data Fig.~\ref{extfig3} were normalized to the maximum value at the largest investigated \SI{130}{nm} NP to compare the trend of their properties.

EEL spectra of hemispherical VO$_2$ particles of various radii were calculated using the commercial software COMSOL Multiphysics, utilizing classical dielectric formalism~\cite{Abajo2010}, and following the procedure reported previously~\cite{Konecna2018}. We assumed the nonrecoil approximation, where the electron trajectory is approximated by a straight line element parallel with the $z$ direction, along which we set the current $I=I_\mathrm{0}\mathrm{e}^{\mathrm{i}\omega z / v}$, where $I_\mathrm{0}$ is the current amplitude, $\hbar\omega$ is the energy loss with $\hbar$ the reduced Planck constant, and $v$ is the electron velocity (set to 0.587 of speed of light to match the acceleration voltage \SI{120}{keV}).
The EEL probability can be expressed as
\begin{multline}
\Gamma_\mathrm{EELS}(\omega)\\
= \frac{e}{\pi\hbar \omega}\int_{z_\mathrm{min}}^{z_\mathrm{max}}\mathrm{d}z\,\mathrm{Re}\Big\{ E_z^{\mathrm{ind}}(\mathbf{R}_\mathrm{b},z,\omega)\mathrm{e}^{\mathrm{-i}\omega z / v}\Big\} ,
\end{multline}
where $E_z^{\mathrm{ind}}$ is the $z$-component of the induced electric field, $\mathbf{R}_\mathrm{b}=(x_\mathrm{b},y_\mathrm{b})$ is the electron beam position relative to the centre of the NP, located in the origin, and $e$ is the elementary charge. We integrate over electron trajectory restricted by the simulation domain $z = (z_\mathrm{min},z_\mathrm{max})$. The boundary conditions in the form of perfectly matched layers were imposed, and the simulation domain was kept sufficiently large (always larger than the wavelength considered). The simulation mesh was refined inside the NPs (max size \SI{7.5}{nm}) and coarser for the surrounding domain, for which we considered $\epsilon_\mathrm{r}=1$.

The material response of all simulated NPs is characterized by the dielectric function (Extended Data Fig.~\ref{extfig8}) obtained by spectroscopic ellipsometry from the VO$_2$ thin film deposited by PLD and annealed for \SI{10}{min} at \SI{600}{\celsius} and \SI{15}{sccm} oxygen flow. 

\section*{Extended Data}

\renewcommand\figurename{Extended Data Fig.}

\begin{figure*}
    \centering
    \includegraphics{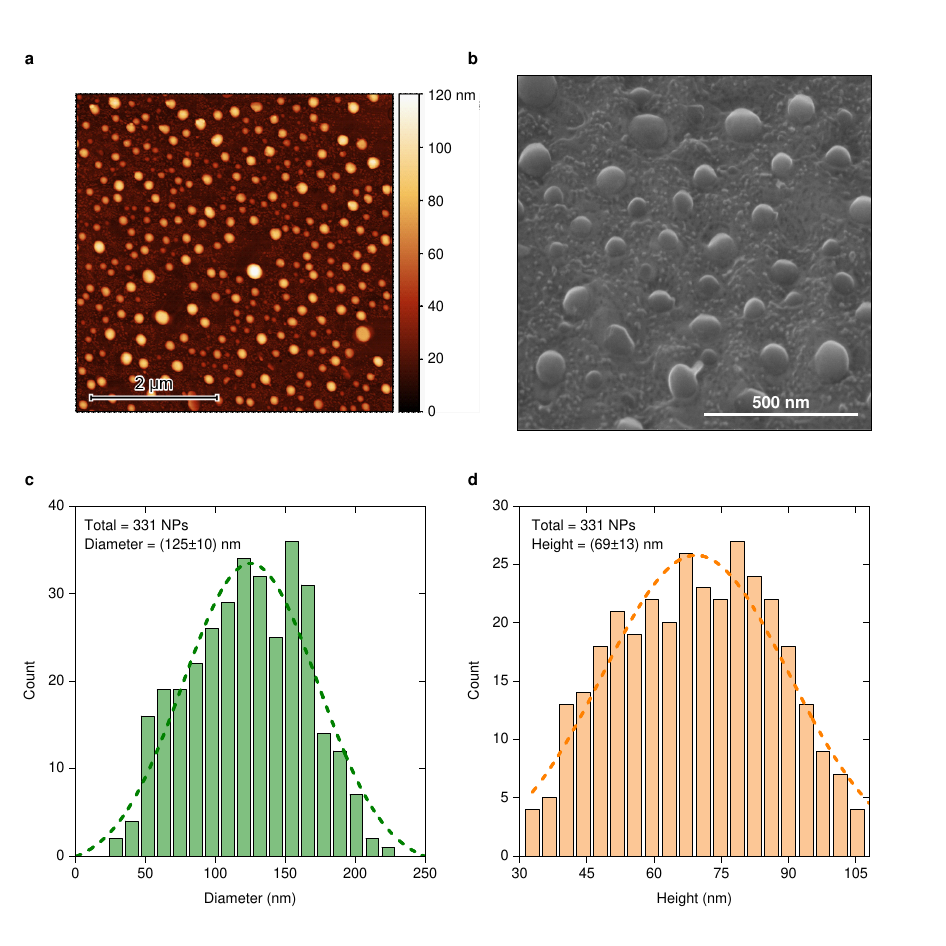}
    \caption{\textbf{Shape and size distribution of VO$_2$ NPs.} \textbf{a}, AFM image and \textbf{b}, SEM micrograph of the VO$_2$ NPs on the fused silica and SiN heating chip membrane, respectively. \textbf{c}, Diameter and \textbf{d}, height distribution of VO$_2$ NPs from the AFM image in (\textbf{a}).}
    \label{extfig1}
\end{figure*}

\begin{figure*}
    \centering
    \includegraphics{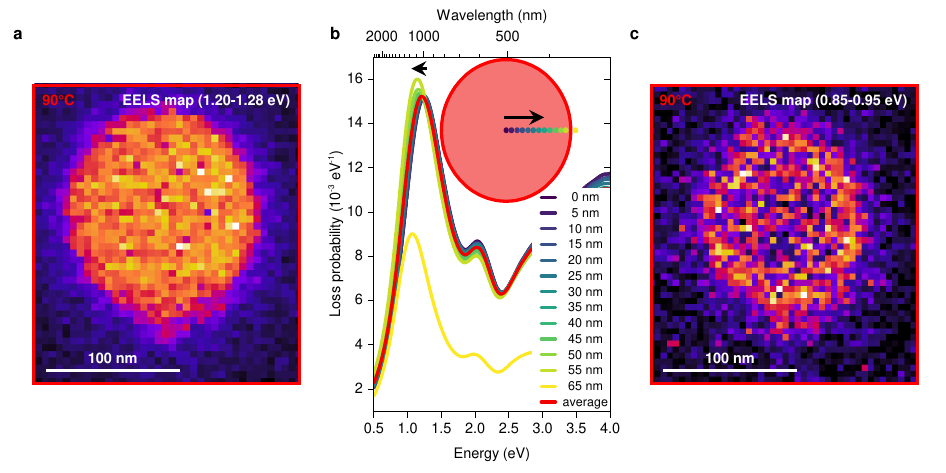}
    \caption{\textbf{Decomposition of the plasmonic peak.} 
    \textbf{a}, EEL intensity map of the investigated NP in Fig.~\ref{fig1}c in the metallic phase, summed between 1.20--\SI{1.28}{eV}. \textbf{b}, Simulated EEL spectra of the metallic spherical VO$_2$ NP with the \SI{130}{nm} diameter in vacuum, illuminated by an electron beam in the listed positions that are also highlighted on the scheme of the NP. \textbf{c}, EEL intensity map of the investigated NP in Fig.~\ref{fig1}c in the metallic phase, summed between 0.85--\SI{0.95}{eV}. The plasmonic peak in Fig.~\ref{fig1}c is formed by two contributions: Predominantly by the volume plasmon, with the homogeneous spatial distribution of the loss probability in (\textbf{a}), and partially by the localized surface plasmon resonance around \SI{1}{eV}, with the EEL probability having a maximum near the boundary of the NP, as shown in (\textbf{b}) and (\textbf{c}).}
    \label{extfig2}
\end{figure*}

\begin{figure*}
    \centering
    \includegraphics{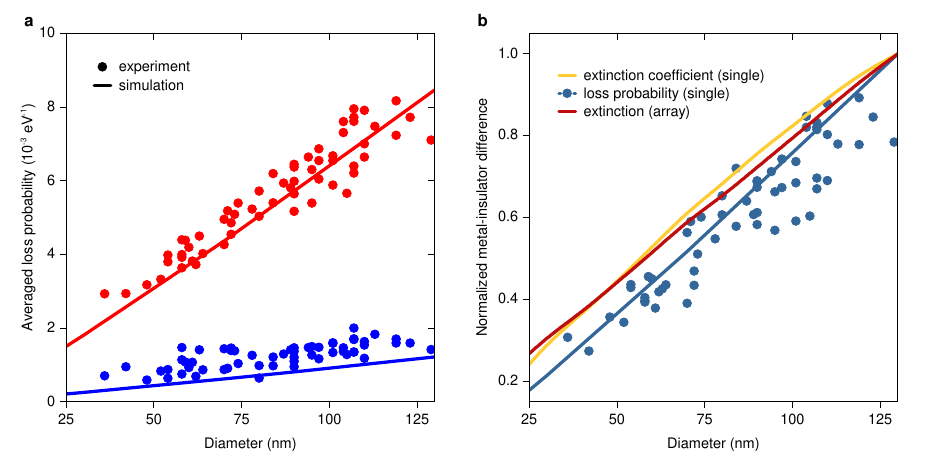}
    \caption{\textbf{Correlation of EEL spectra and far-field optical properties.} \textbf{a}, Experimental and simulated averaged EEL probability of various VO$_2$ NPs as a function of their diameter at the insulator (blue) and metal (red) phase. The averaged EEL probability represents the spatially integrated EEL probability over each NP, and calculated the average value in the region around the resonance between 1.0--\SI{1.5}{eV}. \textbf{b}, Comparison of the simulated optical extinction coefficients and EEL intensities of a single NP and of a periodic array of NPs and experimental EEL intensities in the form of normalized metal-insulator differences of the respective spectra as a function of the NP diameter. Simulated switching contrasts were extracted in the same energy window as in the experiment and normalized to unity. As the calculated optical switching contrasts correspond well to the EEL spectroscopy measurements and all results exhibit approximately a linear dependence on the NP size, we can relate EEL measurements to far-field optical properties and utilize them to investigate possible applications in memories or metasurfaces.}
    \label{extfig3}
\end{figure*}

\begin{figure*}
    \centering
    \includegraphics{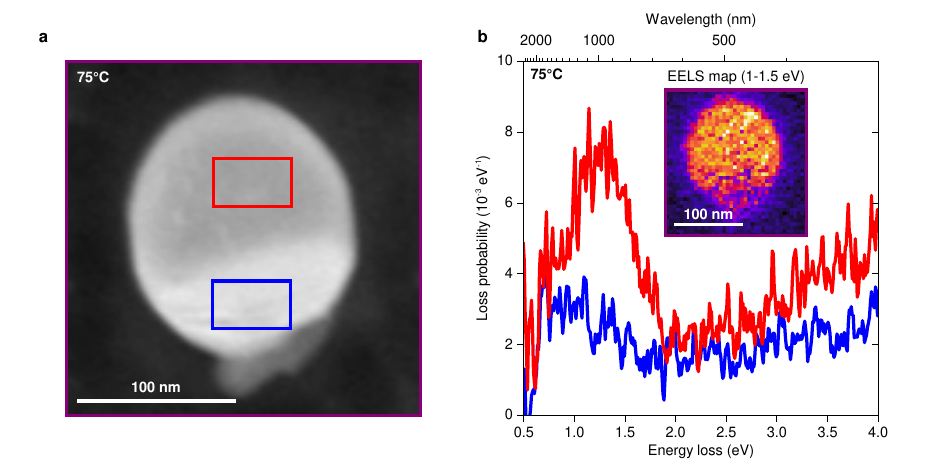}
    \caption{\textbf{EEL spectra of coexisting phases within one NP.} \textbf{a}, ADF image and \textbf{b}, EEL spectra of two coexisting phases within the single VO$_2$ NP recorded at \SI{75}{\celsius}. The inset shows an EEL intensity map of the NP integrated between 1.0--\SI{1.5}{eV}. The blue and red rectangles in (\textbf{a}) highlight the areas from which the spectra were averaged. This figure shows that while the EEL spectrum from the red area implies this region is metallic, the spectrum from the blue area indicates the coexisting insulating phase in that region. Correspondingly, the EEL intensity map of the NP integrated between 1.0--\SI{1.5}{eV} shows low intensity at the bottom of the particle, where this insulating phase is located.}
    \label{extfig4}
\end{figure*}

\begin{figure*}
    \centering
    \includegraphics{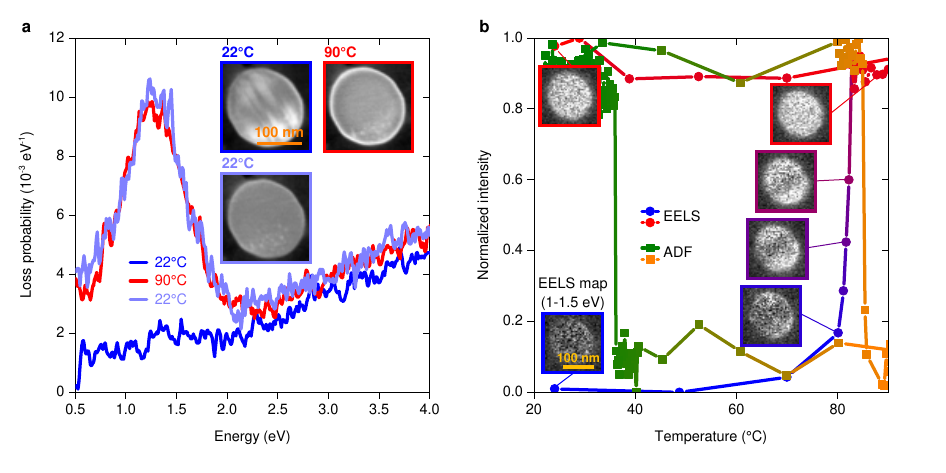}
    \caption{\textbf{Metallic phase conservation by the electron beam.} \textbf{a}, Measured EEL spectra of the VO$_2$ NP with the \SI{150}{nm} diameter obtained at \SI{22}{\celsius} (blue), \SI{90}{\celsius} (red) and again \SI{22}{\celsius} (light blue). Exhibiting the plasmon resonance, the NP remained in the metallic state after cooling back to room temperature. The inset shows ADF images at the listed temperatures. \textbf{b}, Phase transition hysteresis of the normalized EEL intensity averaged between 1.0--\SI{1.5}{eV} and ADF intensity of the NP in (a). The insets show the EEL intensity maps of the NP integrated between 1.0--\SI{1.5}{eV} and measured at the marked temperature positions. The conservation effect was observed for several NPs when illuminated after a certain electron dose. To understand the effect and determine specific doses, a study that is beyond the scope of this paper has to be carried out.}
    \label{extfig5}
\end{figure*}

\begin{figure*}
    \centering
    \includegraphics{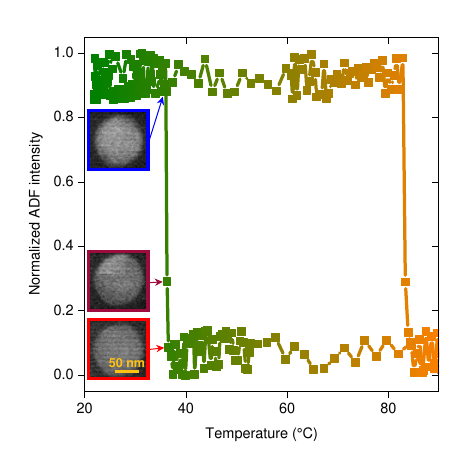}
    \caption{\textbf{Coexisting phases during the MIT.} The phase transition ADF intensity hysteresis of the VO$_2$ NP with the \SI{100}{nm} diameter, extracted from the ADF images. The insets show ADF images during the MIT. The NPs that exhibited coexisting phases during both transitions represent approx. \SI{1}{\percent} of the studied NPs.}
    \label{extfig6}
\end{figure*}

\begin{figure*}
    \centering
    \includegraphics{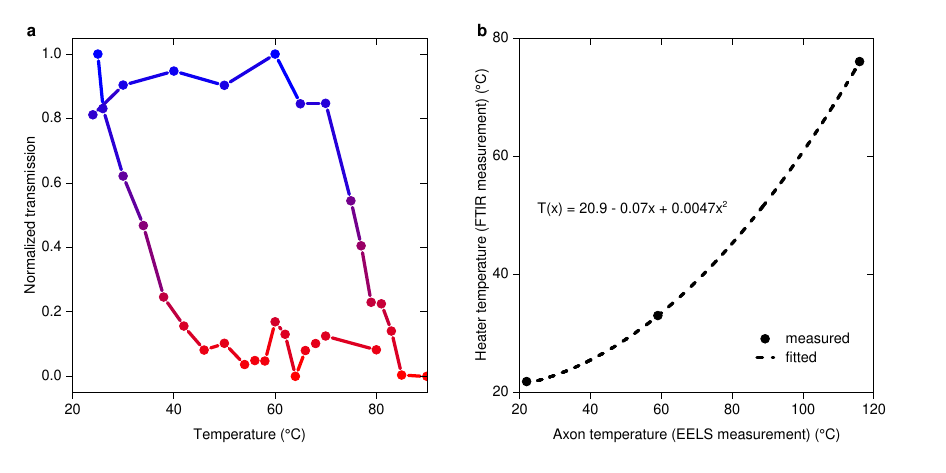}
    \caption{\textbf{SiN heating chip temperature calibration.} \textbf{a}, Phase transition hysteresis of normalized transmission in the near-infrared of VO$_2$ NPs on the SiN heating chip membrane. \textbf{b}, The calibration curve between the temperature displayed in Protochips Inc. software Axon and the temperature displayed by the trusted ex-situ heater. X-coordinates of data points were obtained as the average transition temperatures of hundreds of NPs, processed as in Fig.~\ref{fig3}. Y-coordinates of data points were obtained from the transition temperatures in (\textbf{a}), measured by the ex-situ Fourier transform infrared spectroscopy with the home-built heating stage. The first data point represents an equal room-temperature starting point. Data were fitted by an empirical quadratic function displayed in the graph.}
    \label{extfig7}
\end{figure*}

\begin{figure*}
    \centering
    \includegraphics{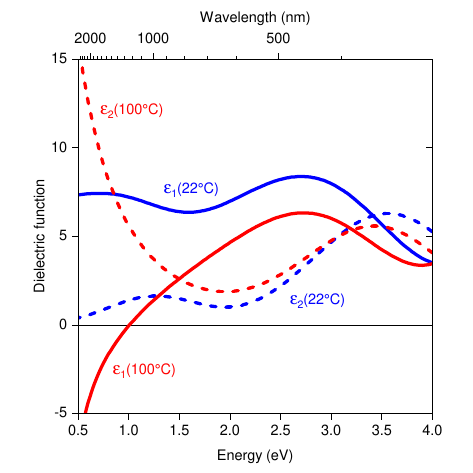}
    \caption{\textbf{Dielectric function.} Real (full line) and imaginary (dashed line) parts of the dielectric function of a \SI{30}{nm} VO$_2$ thin film on the silicon substrate obtained at \SI{22}{\celsius} (blue) and \SI{100}{\celsius} (red) in the insulator and metallic phases, respectively.}
    \label{extfig8}
\end{figure*}

\end{appendices}

\end{document}